\documentclass[twocolumn]{article}
\usepackage{graphicx}
\usepackage{amsmath}
\usepackage{authblk}
\usepackage{amsfonts}
\usepackage{booktabs}
\usepackage{enumerate}
\usepackage{cite}
\usepackage{hyperref}    
\usepackage{siunitx}
\usepackage{url}
\usepackage{braket}
\usepackage{pgfplots}
\usepackage{tikz}

\title{Quantum annealer accelerates the variational quantum eigensolver in a triple-hybrid algorithm}
\author{Manpreet Singh Jattana}
\affil{Modular Supercomputing and Quantum Computing, Goethe University Frankfurt, Kettenhofweg 139, 60325 Frankfurt am Main, Germany}
\affil{jattana@em.uni-frankfurt.de}

\begin{document}

\twocolumn[
\begin{@twocolumnfalse}
	\maketitle
	\begin{abstract}
Hybrid algorithms that combine quantum and classical resources have become commonplace in quantum computing. The variational quantum eigensolver (VQE) is routinely used to solve prototype problems. Currently, hybrid algorithms use no more than one kind of quantum computer connected to a classical computer.  In this work, a novel triple-hybrid algorithm combines the effective use of a classical computer, a gate-based quantum computer, and a quantum annealer. The solution of a graph coloring problem found using a quantum annealer reduces the resources needed from a gate-based quantum computer to accelerate VQE by allowing simultaneous measurements within commuting groups of Pauli operators. 
We experimentally validate our algorithm by evaluating the ground state energy of H$_2$ using different IBM Q devices and the DWave Advantage system requiring only half the resources of standard VQE. Other larger problems we consider exhibit even more significant VQE acceleration.
Several examples of algorithms are provided to further motivate a new field of multi-hybrid algorithms that leverage different kinds of quantum computers to gain performance improvements.
\end{abstract}\quad
\end{@twocolumnfalse}
]

\section{Introduction}
Two major types of quantum computers are currently under research and development \cite{NationalAcademiesofSciencesEngineering2019}. These are gate-based quantum computers \cite{cit22ekey,5088164,GYONGYOSI201951} and quantum annealers \cite{Finnila1994,Harris2010,King2021,Ohzeki2020}. While the annealers are primarily used for specific problems, gate-based quantum computers are universal machines \cite{9781107002173}. 

Many quantum computing algorithms of the current generation are hybrid. These algorithms combine the use of quantum-classical resources to execute tasks. They were designed to ameliorate a challenging problem faced when using current-generation gate-based quantum computers: noise. The variational quantum eigensolver \cite{Peruzzo2014, McClean2016, Wang2019, Colless2017,rev123} and the quantum approximate optimisation algorithm \cite{Farhi2014, PhysRevX.10.021067, dw} are early examples of such algorithms. 

Invented out of necessity, variational algorithms touch upon a vital aspect of computing, emphasising the use of different technologies to undertake specific subtasks of an algorithm more efficiently. Although these hybrid algorithms combine quantum and classical computing, they fail to fully utilize the two primary directions in which the field of quantum computing has emerged. Thus, hybrid algorithms either combine a gate-based quantum computer with a classical computer or a quantum annealer with a classical computer. 

In this paper, we contribute to filling the gap by proposing and implementing a hybrid algorithm combining all three. We believe that this opens a new direction of research that takes advantage of each technology's unique features to accelerate quantum algorithms. We demonstrate an algorithm whose subtasks are divided among a gate-based quantum computer, a quantum annealer, and a classical computer. Such an algorithm works faster than a standard algorithm whose subtasks are divided among a gate-based quantum computer and a classical computer. We execute the new algorithm on several actual quantum computers for each subtask. 

The rest of the paper is structured as follows. In section~\ref{sec112}, we briefly categorize hybrid algorithms to highlight how our work goes beyond existing works. In section~\ref{secmeth}, we introduce the problem we intend to solve using our algorithm, describe the ingredients necessary to formulate it, and explain the working of the triple-hybrid algorithm. In section~\ref{sec3}, we demonstrate our algorithm on a small prototype problem and show additional results for different problems using the quantum annealer. In section~\ref{sec5}, we discuss and conclude the work.
\section{Hybrid algorithms}\label{sec112}
Hybrid quantum-classical algorithms can be categorized into two based on the hardware or algorithm specifications. 

\subsection{Hardware level}
Hybrid algorithms at the hardware level can be categorized based on the number of different types of technologies used within one algorithm:
\begin{enumerate}[I.]
\item \textbf{Double-hybrid}: combining classical computers with one type of quantum computing technology.
\item \textbf{Triple-hybrid}: combining classical computers with two types of quantum computing technologies.
\item \textbf{Multi-hybrid}: combining classical computers with at least three types of quantum computing technologies.
\end{enumerate}

It is possible to have several different units or parts of the same type of technology being used within the algorithm. Examples include parallel quantum annealing \cite{1038,cit222y} and parallel gate-based quantum computing \cite{Niu2023enablingmulti,3352460.3358287,Mineh_2023,9749894}. However, we do not increase the hybrid number where no new technology is introduced in the algorithm.

\subsection{Algorithm level}
Let us consider a double-hybrid case for simplicity but without loss of generality. Three types of situations exist between two hardware technologies that can work in a hybrid setup:
\begin{enumerate}
\item \textbf{Iterative}: when both systems work alternatively such that a loop is established between both with at least two iterations. A bidirectional input-output relationship exists.
\item \textbf{Sequential}: when both systems work one after another sequentially but no loop is established. A unidirectional input-output relationship exists.
\item \textbf{Separative}: when both systems work concurrently but without requiring any significant interaction between one another. No input-output relationship exists.
\end{enumerate}
Examples of \textit{iterative} hybrid algorithms include the variational quantum eigensolver and the quantum approximate optimisation algorithm \cite{Peruzzo2014, McClean2016, Wang2019, Colless2017,rev123,Farhi2014, PhysRevX.10.021067,dw}. 
Solving optimization problems \cite{10.00005,lodewijks2020mapping} on a quantum annealer using forward annealing are examples of \textit{sequential} hybrid algorithms. An example of a \textit{separative} algorithm is when a quantum computer and a classical computer separately try to solve the same problem at the same time and the output of a faster computer or computer with better quality solution is finally chosen while the other is disregarded.

\subsection{Examples}
Below, we mention four potential examples of different types of triple-hybrid algorithms to illustrate how new forefronts of research open up and experimentally demonstrate one of them. 

Example 1: A problem when creating quantum circuits is approximating $\exp (-i t H)$ using $U(t)$, where $H$ is a Hamiltonian in the Pauli basis. The circuit depth can be significantly reduced if commuting terms in $H$ can be found \cite{Raeisi_2012,042303}. Finding the least number of groups in which all terms in a group commute is a hard problem \cite{gokhale2019minimizing}. A quantum annealer can be used to find these groups (see section \ref{secmeth}), and the output is used to create a shorter depth quantum circuit for a gate-based quantum computer instructed by a classical computer. 

Example 2: A job scheduling problem solved on a quantum annealer \cite{venturelli2016quantum} to find an optimal schedule for multiple jobs that require the combined use of a gate-based and classical computers, e.g. variational quantum eigensolver.

Example 3: An important problem in the variational quantum eigensolver is the measurement of all the terms of a Hamiltonian to calculate the expectation value for given variational parameters using the least amount of measurements. These terms can scale as $O(N^4)$ for $N$ qubits \cite{90.022305}. A potential solution is to simultaneously measure commuting terms \cite{jena2019pauli,10.1063}. A quantum annealer can be used to implement this solution in a sequential-hybrid algorithm.

We demonstrate in this work a triple-hybrid algorithm for the above example 3 where a classical computer and quantum annealer are connected as \textit{sequential} hybrid and the classical computer and gate-based quantum computer contribute as \textit{iterative} hybrid. This quantum algorithm can be termed a sequential-iterative triple-hybrid.

Example 4: One can modify example 3 such that the quantum annealer is not used in the forward annealing but reverse annealing \cite{rannea1,022314,arxiv.2303.13748,Venturelli_2019,PhysRevA.100.052321} setup where a feedback loop is established between the classical computer and the quantum annealer. Such a quantum algorithm will be termed an iterative-iterative triple-hybrid. Additionally, standalone reverse quantum annealing could potentially take as input and improve the output of gate-based quantum computers.

\section{Methods} \label{secmeth}

\begin{figure*}
\includegraphics[scale=.87]{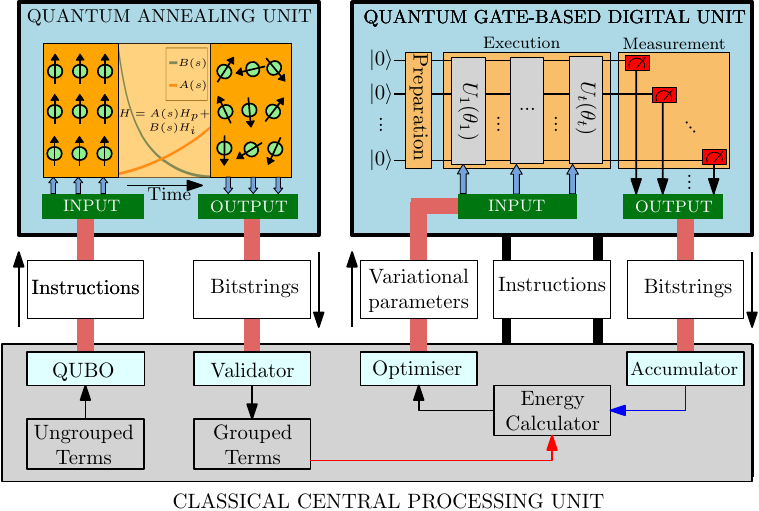}
	\centering
	\caption{Triple-hybrid algorithm combining a quantum annealing unit (QAU), a quantum gate-based digital unit (QDU), and a classical central processing unit (CPU). Ungrouped terms are grouped using a CPU-QAU combination. Grouped terms help reduce the QDU resource consumption by allowing simultaneous measurement of commuting terms.   \label{fchart}}
\end{figure*}

The goal of this section is to introduce the necessary tools to find the ground state energy of a given Hamiltonian using the least amount of measurements by finding the least number of groups of commuting terms. We focus on the problem of finding the ground state energy of a given $N$-qubit Hamiltonian having $m$ terms,
\begin{equation}
 H = \sum_{j=1}^{m} c_j h_j^{\otimes N}, \label{123123}
\end{equation}
where $h_j^{\otimes N} = \otimes_{l=1}^{N} \hat{P}_{jl}$, $\hat{P} \in \{X, Y, Z,I \} $, $X=\sigma^x,Y=\sigma^y,Z=\sigma^z$ are the Pauli matrices, and $I$ is the identity matrix. Finding the ground state energy of problems written in the form of Eq.~(\ref{123123}) is known to be classically difficult in general. One reason for that is the exponential scaling of the classical memory required to store the quantum statevector made up of complex numbers. It is a native quantum problem. Therefore, quantum computers are better candidates.

\subsection{Double-hybrid algorithm}
To the best of the author's knowledge, all current methods to find the ground state of Eq.~(\ref{123123}) using variational quantum algorithms use a double-hybrid setup. This setup includes a gate-based quantum computer and a classical computer working together \cite{rev123}. Greedy algorithms running on classical computers have been previously used to group commuting terms \cite{jena2019pauli,10.1063,izmaylov2019unitary}. However, they rely on heuristics prone to finding a local optimum. They can also incur impractical classical costs \cite{gokhale2019minimizing}. Furthermore, none of them try to exploit the use of a quantum annealer, which can be particularly useful for the graph coloring problem \cite{cite111key,kwok2020graph,Kole2022}.

\subsection{Triple-hybrid algorithm}

We go beyond the existing double-hybrid quantum algorithms and use a triple-hybrid approach as follows.
Figure \ref{fchart} shows how to find the ground state energy using two different quantum computers and a classical computer. The problem Hamiltonian (see Eq.~(\ref{123123})) is input as a string representation to the classical central processing unit (CPU). This information is stored in the box labelled ``Ungrouped terms". These ungrouped terms go through several levels of processing inside the CPU to convert them into a quadratic unconstrained binary optimisation (QUBO) problem, which is sent to the quantum annealing unit (QAU) along with other instructions. These instructions can include the annealing time, qubit mapping, chain strength, number of samples, etc., and are controlled by the user. The QAU tries to solve the QUBO problem and outputs bitstrings that are processed in the CPU. The CPU checks the validity of the solutions obtained from the QAU. One valid solution is then used to group commuting terms stored on a CPU illustrated by the box labelled ``Grouped terms". These are used for the next stage of computation.

The next stage is a combination of a CPU and a quantum gate-based digital unit (QDU). This particular setup has been widely used in the context of a variational quantum eigensolver \cite{PhysRevApplied.19.024047,willsch2022hybrid}. The computation is started by giving initial parameters to a classical optimisation algorithm, optimiser in short, running on the CPU. The CPU then passes these parameters embedded into a quantum circuit to the QDU. The CPU also passes additional instructions to the QDU; see Ref. \cite{jattana} for an extensive list. The relevant instructions for our purposes are the number of samples and the measurement basis for a group of terms in the problem Hamiltonian. CPU instructs the QDU to produce bitstrings for each of the groups created with the help of the QAU.

The QDU executes the circuit for the given set of parameters and instructions and produces bitstrings as an output. These bitstrings are accumulated in the CPU using the ``Accumulator" as a storage unit. Once sufficient bitstrings are accumulated, they are used to calculate the energy for all the terms in all the groups to obtain the energy of the Hamiltonian. This energy is then the scalar quantity minimised by the optimiser in subsequent iterations. The algorithm stops when the optimiser can no longer further minimise the energy or a certain preordained number of iterations have been reached.

In the following sections, we go through the details of how the above algorithm is realised. We discuss the commutativity of the terms given in Eq.~(\ref{123123}) and how commuting groups are formed using the graph colouring problem. We illustrate the algorithm step-by-step using simple examples.

\subsection{Commutation}
We analyse the commutativity between two terms of a Hamiltonian. This analysis will lead to the formulation of a graph coloring problem. Two terms $h_i$ and $h_j$ commute when
\begin{equation}
	[h_i,h_j] = h_ih_j-h_jh_i = 0.
\end{equation}
In other words, two terms commute if their commutator, given by the [ ] symbol, is zero. Note that any two terms commute trivially for $i=j$. The commutativity of terms in a Hamiltonian can be formulated in two different ways, namely, qubit-wise commutativity (QWC) and general commutativity (GC) \cite{Raeisi_2012, gokhale2019minimizing}.

In QWC, two terms are said to commute if the Pauli terms corresponding to the same index between two Pauli strings commute with one another. For example, the term $h_1 = X_1X_2$ qubit-wise commutes with $h_2 = X_1I_2$ but not with $h_3 = Y_1Y_2$. We can create a set of terms using QWC such that all the terms in this group qubit-wise commute with every other term, e.g. $\{h_1,h_2\}$. The idea is that all the terms of a group can be measured simultaneously in a QDU, thereby facilitating quicker computation. QWC has been used in experimental demonstrations of small systems on quantum processors \cite{44,45,46}.

In GC, two terms are said to commute if the Pauli terms corresponding to the same index between two Pauli strings fail to qubit-wise commute an even number of times. In this case, for the above example, $h_1$ commutes with $h_2$ and $h_3$, but $h_2$ does not commute with $h_3$. Note that QWC is a limited special case of GC. The benefit of using GC is that it can potentially include more terms in a group compared to using QWC. The drawback of using GC over QWC is that the simultaneous measurement of terms in a GC group is non-trivial, while that of a QWC group is trivial. 

Although methods to measure a GC group have been devised using stabiliser matrices \cite{gokhale2019minimizing}, they are outside the scope of this work. In this work, we will only consider QWC from now on; therefore, it may be that some terms that actually commute (as per GC) are not grouped together. However, we will see that there are significant benefits nonetheless.

\subsection{Graph colouring}
Although we want to maximise the size of each group, the central problem is to minimise the number of groups. Thus, we are looking to solve a \textit{minimum clique cover} problem where each clique is the group of commuting terms. If the minimum clique cover problem is given by a graph $G$, then the complement graph $\bar{G}$ gives the corresponding graph colouring problem \cite{10.1063}. Both these belong to the class of problems considered to be NP-hard in general \cite{kp1972}. The graph colouring problem can be converted to a quadratic unconstrained binary optimisation (QUBO) problem \cite{glover} and solved on a QAU, cf. \cite{kwok2020graph,Chancellor_2019,9259934}. This section briefly explains the steps needed to achieve this objective.

The problem is input as a list of strings, each string of length $N$ corresponding to one term in the Hamiltonian, and converted to an integer format representing the indices of the Pauli terms \cite{jattana}. This format is conveniently manipulated to find out the commutativity of each term with every other. The commutativity information is encoded into a graph as follows. Create a graph with $m$ nodes, where $m$ is the total number of terms in the Hamiltonian. If a given term $h_i$ does not commute with another term $h_j$, i.e. $[h_i,h_j] \neq 0$, add an edge between the vertices indexed $i$ and $j$.

In graph colouring, no two neighbours are allowed to have the same colour. In our problem, no two non-commuting terms can be in the same group. Therefore, by creating edges between indices representing non-commuting terms, we have successfully translated our problem into a graph colouring problem. Solving the graph colouring problem will now yield us the solution to our original problem of finding commuting groups of terms.

\begin{figure}
	\includegraphics[scale=1.3]{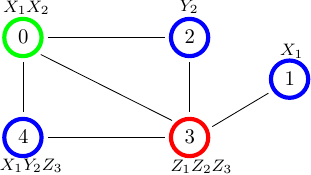}
	\centering
	\caption{The indexed vertices in the graph represent the terms and the edges represent the non-commutativity between the terms for an example Hamiltonian given in Eq.~(\ref{eq:1}). The colouring of the vertices represents one way to colour the graph. \label{fig1label}}
\end{figure}

As an example, consider the following arbitrary Hamiltonian 
\begin{equation}
	H_{\textmd{\footnotesize ex.}} = X_1X_2 + X_1 + Y_2 + Z_1Z_2Z_3 + X_1Y_2Z_3. \label{eq:1}
\end{equation}
The commutativity between the terms is shown in the graph in Fig.~\ref{fig1label}. The terms are indexed in the sequence they appear in Eq.~(\ref{eq:1}) from left to right. The edges between any two terms represent the qubit-wise non-commutativity of those terms. The graph would look different for general non-commutativity. The minimum colours required to colour this graph are 3, which is known as the chromatic number of the graph. Note that the colouring shown in Fig.~\ref{fig1label} is not unique, for example, vertex $1$ could also have the green colour. According to the colouring scheme in Fig.~\ref{fig1label}, we can measure the terms $X_1$, $Y_2$, and $X_1Y_2Z_3$ simultaneously.

\subsection{QUBO formulation}
In this section, we briefly summarise the steps required to convert the graph colouring problem to a QUBO problem. Assume that we wish to colour a graph with exactly $K$-colours. Let $x$ be a binary variable such that $x_{ij}=1$ if vertex $i$ is assigned colour $j$, and $0$ otherwise. The graph colouring has two constraints, namely, (a) each vertex must receive a colour, and (b) connected adjacent vertices are always assigned different colours. 
The constraint (a) is imposed such that
\begin{equation}
	\sum_{j=1}^{K}x_{ij}  = 1 \quad \text{for} \quad i=1,\dots,N, \label{eq4}
\end{equation}
where $N$ is the number of vertices in the graph. Thus, the binary variable is $1$ only for one out of $K$ colours. We observe that $x_{ij} = x_{ij}^2$, since the variable is binary.  The constraint (b) is imposed such that
\begin{equation}
x_{ip}+x_{jp}\leq 1 \quad \text{for} \quad p=1,\dots,K, \label{eq5}
\end{equation}
and for all adjacent vertices $i$ and $j$. The QUBO is expressed by the optimization problem:
\begin{equation}
\text{QUBO: min } \mathbf{x}^t Q \mathbf{x},
\end{equation}
where $Q$ is a square matrix of constants and incorporates all the information about our problem, and $\mathbf{x}$ is a vector of binary decision variables. The steps to convert Eq.~(\ref{eq4}) and Eq.~(\ref{eq5}) into $Q$ are detailed in Ref. \cite{glover}, and include using two different transformations and using single subscripts to impose penalties for violating the problem constraints. The creation of $Q$ can be automated on a CPU for a given graph $G$. We generate the $Q$ matrices in this work using the \textit{qubogen} package \cite{qubogen}, which is based on the graph colouring formulation given in Ref. \cite{glover}.

\section{Results}\label{sec3}
In this section, we show results for two cases. First, we demonstrate a use-case of our algorithm where we find the ground state energy of a small two-qubit Hamiltonian using a QAU-CPU-QDU hybrid algorithm. Second, we take Hamiltonians with different number of qubits and terms, and use both QWC and GC to group the terms into commuting sets using a QAU. 

\begin{table}
		\setlength{\tabcolsep}{4pt} 
\centering{
		\begin{tabular}{c|cccccccc|c|c}
			\toprule					
		&\multicolumn{8}{c|} {Indexed bitstring}	\\
		\cmidrule{2-9}
		\multicolumn{1}{c} {} &\multicolumn{1}{|c} {1}&\multicolumn{1}{c} {2}&\multicolumn{1}{c} {3}&\multicolumn{1}{c} {4}&\multicolumn{1}{c} {5}&\multicolumn{1}{c} {6}&\multicolumn{1}{c} {7}&\multicolumn{1}{c|} {8}& \multicolumn{1}{c|} {Energy}& \multicolumn{1}{c|}{Frequency} \\
			\midrule
		1&	{$0$} & {$1$} &{$0$} &{$1$} &{$0$} &{$1$} &{$1$} &{$0$} & {$-16.0$}  & {$476$}  \\ 
		2&	{$1$} & {$0$} &{$1$} &{$0$} &{$1$} &{$0$} &{$0$} &{$1$} & {$-16.0$}  & {$520$}  \\ 
		3&	{$0$} & {$1$} &{$1$} &{$0$} &{$0$} &{$1$} &{$1$} &{$0$} & {$-12.0$}  & {$1$}  \\ 
		4&	{$1$} & {$0$} &{$1$} &{$0$} &{$0$} &{$1$} &{$0$} &{$1$} & {$-12.0$}  & {$1$}  \\ 
		5&	{$0$} & {$1$} &{$0$} &{$0$} &{$0$} &{$1$} &{$1$} &{$0$} & {$-12.0$}  & {$2$}  \\ 
			\bottomrule
		\end{tabular}%
	}
		\caption{Energy and frequency of occurrence related to the five unique bitstrings obtained from the QAU when sampling a $1000$ times. Energy in this table is a dimensionless quantity.}
	\label{rawdata}%
\end{table}%

\subsection{$\text{H}_2$ molecule}
The algorithm is demonstrated by finding the ground state energy of the $\text{H}_2$ molecule \cite{44} whose Hamiltonian is given by
\begin{equation}
	\begin{split}
 H_{h2} &=  0.011* Z_1Z_2 +0.398* Z_1 \\
 &+ 0.398* Z_2 + 0.181 * X_1X_2. \label{eqh2}
	\end{split}
\end{equation}
By using QWC, the commutativity of each term is computed in relation to every other term. During this process, by using the Python package \textit{networkx} \cite{SciPyProceedings_11}, an edge is added on a null graph between two vertices (representing terms) which do not commute. The final graph looks the same as Fig. \ref{fig1label}, if we remove the vertex $0$ and its edges and assign the term $X_1X_2$ to vertex $3$, $Z_1$ to vertex $1$, $Z_2$ to vertex 2, and $Z_1Z_2$ to vertex 4. The \textit{networkx} package offers an in-built greedy algorithm which could colour the graph using two colours, which is also the chromatic number.

\begin{table}
	\setlength{\tabcolsep}{4pt} 
	\centering{
		\begin{tabular}{c|cc|cc}
			\toprule					
			&\multicolumn{2}{c} {Naive grouping} &\multicolumn{2}{|c|} {QWC grouping}	\\
			\cmidrule{2-5}
			\multicolumn{1}{|c} {Device} &\multicolumn{1}{|c} {Runs}&\multicolumn{1}{c|} {Energy}&\multicolumn{1}{c} {Runs}&\multicolumn{1}{c|} {Energy}\\
			\midrule
			Emulator&	{$4$} & {$-0.192$} &{$2$} &{$-0.192$} \\ 
			IBMQ Jakarta&	{$4$} & {$-0.139$} &{$2$} &{$-0.141$} \\ 
			IBMQ Manila&	{$4$} & {$-0.125$} &{$2$} &{$-0.116$} \\ 
			IBMQ Perth&	{$4$} & {$-0.099$} &{$2$} &{$-0.104$} \\ 
			IBMQ Nairobi&	{$4$} & {$-0.126$} &{$2$} &{$-0.130$} \\ 
			\bottomrule
		\end{tabular}%
	}
	\caption{Ground state energies obtained for different QDUs using naive and QWC grouping. Energy in this table has the Hartree atomic units. The triple-hybrid algorithm needs only half the number of QDU runs using QWC grouping.
}
	\label{rawdatadev}%
\end{table}%

The penalty variable for the QUBO problem was set to $4$ and the number of colours to $2$. The number of qubits required equals the product of the number of colours and vertices. Eight qubits were used for this particular case. The number of samples was set to $1000$. The QAU used was the DWave Advantage System version 5.3 \cite{dwave}. The raw data obtained are tabulated in Table \ref{rawdata}.

The bitstrings obtained from the QAU are processed in a ``Validator" (see Fig. \ref{fchart}) to check if they satisfy all the constraints of the problem. Only the data in the top two rows of Table \ref{rawdata} with energy $-16$ satisfies the constraints. These two bitstrings correspond to colouring the vertex corresponding to the $X_1X_2$ term differently from the other three. From the first bitstring, we make the groups \{$X_1X_2$\} and \{$Z_1$, $Z_2$, $Z_1Z_2$\}. We note that sampling the QAU automatically gives more than one possible valid solution (if it exists). Invalid solutions also appear in the samples, in this case with a very low frequency of occurrence.

The next step is to prepare a parametrised ansatz on the QDU. For our problem, we take the ansatz used in Ref. \cite{PhysRevX.6.031007} given by 
\begin{equation}
 \ket{\psi(\theta)} = e^{-i\theta X_1Y_2} \ket{01}.\label{eqxy}
\end{equation}
The energy for an arbitrary value of $\theta \in [0,2\pi]$ is given by $E(\theta) = \braket{\psi(\theta)|H_{h2}|\psi(\theta)}$, where $\ket{\psi}$ is assumed to be normalised. The ground state energy is then given by 
\begin{equation}
E_0 = \braket{\psi(\Theta)|H_{h2}|\psi(\Theta)}
\end{equation}
where $\Theta$ corresponds to that value of $\theta$ where the global minimum is located. For the Hamiltonian given in Eq. (\ref{eqh2}) and the ansatz given in Eq. (\ref{eqxy}), the global minimum is located at $\Theta = \pi/2$. For the sake of our demonstration, we forgo the optimisation process to find $\Theta = \pi/2$ but use it directly to calculate the ground state energy on the IBM-Q QDUs Jakarta \cite{dev:jakarta}, Manila \cite{dev:manila}, Perth \cite{dev:perth}, Nairobi \cite{dev:nairobi}, and on an ideal emulator. 

Using a naive method, all four terms are measured separately, thus requiring four experiments. The QWC grouping method requires only two experiments, one in the $Z_1Z_2$ basis and the other in the $X_1X_2$ basis. A basis is prepared using appropriate rotation gates \cite{jattana}. When using QWC, the energy for the $Z_1$ and $Z_2$ terms is calculated using the bitstrings obtained for $Z_1Z_2$. We sample the QDU in each experiment $2^{13}$ times. The precision of the final energy can be improved by sampling more times.

The energies obtained from the QDU are tabulated in Table \ref{rawdatadev}. The results from the ideal emulator can be used as a benchmark to compare the results from the IBM Q devices. We used four different devices and found the energies using (1) the naive grouping, where all terms are measured separately, and (2) the QWC grouping. We found that the energies in both cases were close to one another, as expected. The benefit of the QWC grouping method, in this case, is that it needs only half the computational time than the naive grouping method since only two runs are required. We note that the device performances varied, and none of the devices could find the ground state energy which matches the accuracy of the emulator. This reflects the fact that the current generation devices are in the noisy intermediate scale quantum (NISQ) era. Minor improvements in the energy may have been possible if the circuit parameter optimisation was performed individually for each device; however, this was not relevant for the demonstration of our setup.

In this section, it was shown that a QAU, QDU, and a CPU can be used together in a hybrid setup to execute an algorithm. The triple-hybrid algorithm was executed on currently available actual quantum computers. The usage of one type of quantum technology, i.e., QAU, helped reduce the resources needed from another technology, i.e., QDU, by a factor of two in our example. In the next section, we show how the reduction in resources can be even larger for larger problems.

\subsection{Speed-up using QAU}
We now shift our focus to solving the graph colouring problem for larger problems on the QAU. We consider those Hamiltonians that have either already been used as prototypes on current generation QDUs or have the potential to be used in the near future. These include small molecules, the Hubbard model, and the Heisenberg model \cite{gokhale2019minimizing, fphy.2022.907160}. We use the OpenFermion package \cite{McClean_2020} to generate Hamiltonians for the Hubbard model. The QAUs used were the DWave Advantage System version 5.3 and DWave Advantage2 prototype1.1 \cite{dwave}.

\begin{table}
{
		\begin{tabular}{c|c|cc|cc}
			\cmidrule{3-6}
						\cmidrule{3-6}
			\multicolumn{1}{c} {}	& \multicolumn{1}{c|} {}& \multicolumn{2}{c|}{Qubit-wise} & \multicolumn{2}{c|}{General}\\
			\cmidrule{3-6}
			\multicolumn{1}{c|} {Hamiltonian   } & \multicolumn{1}{c|}{Terms} 	& \multicolumn{1}{c|} {\small G} & \multicolumn{1}{c|} {\small QAU} & \multicolumn{1}{c|} {\small  G} & \multicolumn{1}{c|} {\small  QAU}\\
			\midrule
			{$\text{H}_2$ \cite{44}} &4(3)& 2& 2  &  $2$&  $2$   \\ 
			{$\text{LiH}$ \cite{44}} &30 &  3& 3  & 3 &3     \\ 
			{$\text{H}_2$ (BK) \cite{PhysRevX.8.031022}} &5 &3&3&2&2 \\ 
			{$\text{H}_2$ (JW) \cite{PhysRevX.8.031022}} &14(8) &5 &5 &$2$ &$2$ \\ 
			{$\text{Hubb.} \quad 2\times2$ } & 16(28)   & 8  & -& 3& 3    \\
			{$\text{Hubb.} \quad 1\times3$ } &   21 &9 & - & 4 &3    \\ 
			{$\text{Heis.} \quad 1\times20$ } &   60 & 3  & 3 &2 & 2  \\
			{$\text{Heis.} \quad 3\times3$ } & 36  &3   & 3 &4&3   \\
			\bottomrule
		\end{tabular}%
	}	\caption{Term grouping results from Greedy and QAU solvers. A speed-up for the VQE is seen for all cases. Except for three cases, the same number of terms were used for QWC and GC. Parenthesis is for GC terms. G and QAU stand for Greedy algorithm and quantum annealing unit, respectively. QAU was not able to solve the qubit-wise case for the Hubbard model lattices.  }
	\label{tab:addlabel}%
\end{table}%

The Hamiltonians from quantum chemistry and the Hubbard model are hard problems to group into commuting terms. They offer the benefit that the number of terms in a molecular Hamiltonian can be increased or decreased by selecting a smaller or a larger basis, thus, changing the problem size and difficulty as necessary. Alternatively, the Heisenberg model in one dimension is trivially grouped into three groups using QWC, and serves as a benchmarking problem for the QAU as well as the classical greedy solvers. 

Sometimes a few terms appear in a Hamiltonian that commute with all other terms. In graph colouring, these terms are not part of the connected graph because their commutator is zero with all the terms. These terms can be grouped into any of the groups. We simplify our problem by removing these terms from the Hamiltonian in the examples below. Additionally, when using the Jordan-Wigner transformation, there are often terms with products of only $Z$ operators. Since terms with only such products trivially commute with each other, one can remove them from the terms and make one group out of them. 

Table \ref{tab:addlabel} shows the results of grouping commuting terms for different Hamiltonians. The number of terms is listed for each case. Note that the $Z_1Z_2$ term in  Eq.~(\ref{eqh2}) commutes with all terms using GC; therefore, it was removed when using GC. The number of terms for $\text{LiH}$ molecule was too large for the QAUs; therefore, we selected only a subset from Ref. \cite{44}. In the case of the Hubbard model on a $2\times 2$ lattice, which had $28$ terms, we removed $8$ $Z$-only terms for the QWC case. In some cases, due to the removal of the terms, not the same number of terms was used for QWC or GC, and in this case the number of terms in the parenthesis is the number of terms for the GC case.

We first run the greedy algorithm to find the minimum number of colours needed to colour the graph. We then task the QAU to colour the same graph using this number of colours. The molecular and Hubbard Hamiltonians need to be converted to the spin Hamiltonians using either the parity, Jordan-Wigner (JW), or the Bravyi-Kitaev (BK) transformations \cite{1063}. This is one of the reasons why the $\text{H}_2$ Hamiltonian has a different number of terms. We observe that the QAU is able to correctly colour the graph in all cases except for QWC of Hubbard models.

\begin{figure}
\includegraphics{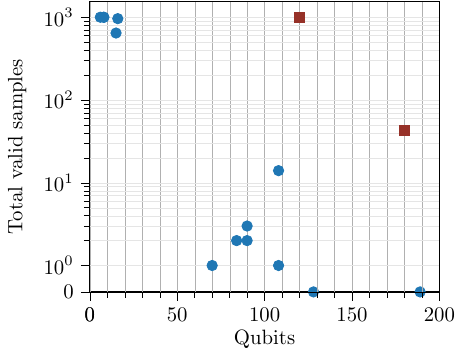}
\caption{The total number of valid solutions obtained when sampling $10^3$ times the QUBO formulation of different problem Hamiltonians given in Table \ref{tab:addlabel} as a function of the number of qubits needed to implement the problem on a QAU. The two outliers in the trend are shown in red squares and belong to the case of Heisenberg model. }\label{figvalid}
\end{figure}

Figure~\ref{figvalid} plots the number of valid solution samples obtained when sampling a QAU $10^3$ times for each of the problems given in Table \ref{tab:addlabel}. The x-axis enumerates the number of qubits needed to implement the problem on the QAU, which is the product of the number of terms and colours for a given problem. We observe the general trend that the frequency of obtaining a valid sample reduces as we increase the number of qubits, eventually going to zero. In two cases, no valid solution was found (see Table \ref{tab:addlabel}). There are two outliers in the data at qubits $120$ and $180$ which correspond to GC and QWC of the Heisenberg model lattices of size $1\times 20$, respectively. These are the benchmarking problems whose solutions are known but offer relatively large problems sizes to test a QAU. However, they too follow the general trend that the valid samples decrease with increasing number of qubits.

In summary, in those cases where we obtained a valid solution, we expect a speed-up in VQE due to the grouping of the terms using the QAU proportional to the number of total terms divided by the number of groups. In Table \ref{tab:addlabel}, this speed-up ranges between a factor of $1.5$x and $30$x.

\section{Discussion and conclusion}\label{sec5}
The greedy algorithm is a useful tool to find optimal colouring for the prototype problems we demonstrated and some other small scale problems we tested. However, being a heuristic, it will likely fail to find the colouring with the least number of colours when the problem size is increased. Furthermore, the greedy algorithm employs different strategies to find the solution. For example, we used the \textit{largest first} strategy \cite{10.10931.85} with a runtime of order $O(m+e)$ where $m$ and $e$ are the number of vertices and edges in $G$, respectively \cite{10.1063}. This strategy failed to find the optimal colouring for one of the problems we tested, namely, the GC case of $3\times 3$ Heisenberg lattice (see Table \ref{tab:addlabel}). Interestingly, the QAU was able to find the correct solution using three instead of four colours.

Due to the way in which a QUBO is formulated, a QAU always needs the number of colours needed as an input. Finding the lowest number of colours needed is itself NP-hard in general. Therefore, it is possible to envision a hybrid algorithm which starts with the solution obtained from a greedy algorithm and inputs it to an annealer which then tries to improve the solution by using fewer colours. Such a setup would be another example of an iterative QAU-CPU hybrid algorithm. This can be explored in future works.

While the simultaneous measurement of commuting groups reduces the total runtime of the VQE, solving the graph coloring problem on the quantum annealer also consumes computing time. This adds to the total runtime of the triple-hybrid algorithm. However, obtaining samples for the VQE required significantly more runtime on QDUs than solving graph coloring on QAUs, which must run only a fixed number of times initially. Future works should study the quantitative time benefit obtained and ways to optimise it.

The current setup focusing on accelerating VQE by reducing the number of measurements remains independent of other methods to accelerate VQE \cite{6.013205,Wang2019} or improved embedding techniques \cite{Chancellor_2019}. These can be suitably integrated into the triple-hybrid setup. Classical post-processing methods \cite{Cao_2023,citekeyem}, among others, can potentially further increase the benefits obtained.

Given that the quantum computing technologies are under development and unlike classical computers, have not attained maturity, there are limitations to the extent to which they can be used together. An important metric to look for is the quantum technology readiness levels. On this metric, the current generation QAUs are ahead of the QDUs \cite{909654}. The technology readiness for our algorithm in the decreasing order is CPU $>$ QAU $>$ QDU. Therefore, we are currently restricted by the QDUs when trying to solve larger problems using the proposed algorithm. This can be expected to change as the technologies mature in the future.

In conclusion, we demonstrated a novel triple-hybrid algorithm on real quantum hardware that used one type of quantum computer to reduce resource consumption and thereby accelerate another type of quantum computer.

\section{Acknowledgments}

I acknowledge the use of IBM Quantum services for this work. The views expressed are those of the author, and do not reflect the official policy or position of IBM or the IBM Quantum team. The author gratefully acknowledges the Jülich Supercomputing Centre (https://www.fz-juelich.de/ias/jsc) for funding this project by providing computing time on the D-Wave Advantage™ System JUPSI through the Jülich UNified Infrastructure for Quantum computing (JUNIQ). The author acknowledges conceptual inputs from Thomas Lippert and Bernhard Frohwitter through discussions. Special thanks are extended to Fengping Jin and Vrinda Mehta for their valuable feedback on writing this paper.

\bibliographystyle{prsty} 
\bibliography{reference} 

\begin{thebibliography}{10}

\bibitem{NationalAcademiesofSciencesEngineering2019}
{National Academies of Sciences, Engineering} and Medicine,  in {\em {Quantum
  Computing: Progress and Prospects}}, edited by E. Grumbling and M. Horowitz
  (The National Academies Press, Washington, DC, 2019).

\bibitem{cit22ekey}
F. Arute, K. Arya, R. Babbush, D. Bacon, J.~C. Bardin, R. Barends, R. Biswas,
  S. Boixo, F.~G. S.~L. Brandao, D.~A. Buell, B. Burkett, Y. Chen, Z. Chen, B.
  Chiaro, R. Collins, W. Courtney, A. Dunsworth, E. Farhi, B. Foxen, A. Fowler,
  C. Gidney, M. Giustina, R. Graff, K. Guerin, S. Habegger, M.~P. Harrigan,
  M.~J. Hartmann, A. Ho, M. Hoffmann, T. Huang, T.~S. Humble, S.~V. Isakov, E.
  Jeffrey, Z. Jiang, D. Kafri, K. Kechedzhi, J. Kelly, P.~V. Klimov, S. Knysh,
  A. Korotkov, F. Kostritsa, D. Landhuis, M. Lindmark, E. Lucero, D. Lyakh, S.
  Mandr{\`a}, J.~R. McClean, M. McEwen, A. Megrant, X. Mi, K. Michielsen, M.
  Mohseni, J. Mutus, O. Naaman, M. Neeley, C. Neill, M.~Y. Niu, E. Ostby, A.
  Petukhov, J.~C. Platt, C. Quintana, E.~G. Rieffel, P. Roushan, N.~C. Rubin,
  D. Sank, K.~J. Satzinger, V. Smelyanskiy, K.~J. Sung, M.~D. Trevithick, A.
  Vainsencher, B. Villalonga, T. White, Z.~J. Yao, P. Yeh, A. Zalcman, H.
  Neven, and J.~M. Martinis, Nature {\bf 574},  505  (2019).

\bibitem{5088164}
C.~D. Bruzewicz, J. Chiaverini, R. McConnell, and J.~M. Sage, Applied Physics
  Reviews {\bf 6},  021314  (2019).

\bibitem{GYONGYOSI201951}
L. Gyongyosi and S. Imre, Computer Science Review {\bf 31},  51  (2019).

\bibitem{Finnila1994}
A.~B. Finnila, M.~A. Gomez, C. Sebenik, C. Stenson, and J.~D. Doll, Chemical
  Physics Letters {\bf 219},  343  (1994).

\bibitem{Harris2010}
R. Harris, M.~W. Johnson, T. Lanting, A.~J. Berkley, J. Johansson, P. Bunyk, E.
  Tolkacheva, E. Ladizinsky, N. Ladizinsky, T. Oh, F. Cioata, I. Perminov, P.
  Spear, C. Enderud, C. Rich, S. Uchaikin, M.~C. Thom, E.~M. Chapple, J. Wang,
  B. Wilson, M.~H.~S. Amin, N. Dickson, K. Karimi, B. Macready, C.~J.~S.
  Truncik, and G. Rose, Phys. Rev. B {\bf 82},  024511  (2010).

\bibitem{King2021}
A.~D. King, C. Nisoli, E.~D. Dahl, G. Poulin-Lamarre, and A. Lopez-Bezanilla,
  Science {\bf 373},    (2021).

\bibitem{Ohzeki2020}
M. Ohzeki, Scientific Reports {\bf 10},    (2020).

\bibitem{9781107002173}
M.~A. Nielsen and I.~L. Chuang, {\em Quantum Computation and Quantum
  Information: 10th Anniversary Edition} (Cambridge University Press,
  Cambridge, 2011).

\bibitem{Peruzzo2014}
A. Peruzzo, J. McClean, P. Shadbolt, M.~H. Yung, X.~Q. Zhou, P.~J. Love, A.
  Aspuru-Guzik, and J.~L. O'Brien, Nature Communications {\bf 5},  4213
  (2014).

\bibitem{McClean2016}
J.~R. McClean, J. Romero, R. Babbush, and A. Aspuru-Guzik, New Journal of
  Physics {\bf 18},  23023  (2016).

\bibitem{Wang2019}
D. Wang, O. Higgott, and S. Brierley, Phys. Rev. Lett. {\bf 122},  140504
  (2019).

\bibitem{Colless2017}
J.~I. Colless, V.~V. Ramasesh, D. Dahlen, M.~S. Blok, M.~E. Kimchi-Schwartz,
  J.~R. McClean, J. Carter, W.~A. de~Jong, and I. Siddiqi, Phys. Rev. X {\bf
  8},  011021  (2018).

\bibitem{rev123}
M. Cerezo, A. Arrasmith, R. Babbush, S.~C. Benjamin, S. Endo, K. Fujii, J.~R.
  McClean, K. Mitarai, X. Yuan, L. Cincio, and P.~J. Coles, Nature Reviews
  Physics {\bf 3},  625  (2021).

\bibitem{Farhi2014}
E. Farhi, J. Goldstone, and S. Gutmann, A Quantum Approximate Optimization
  Algorithm, arXiv:~1411.4028, 2014.

\bibitem{PhysRevX.10.021067}
L. Zhou, S.-T. Wang, S. Choi, H. Pichler, and M.~D. Lukin, Phys. Rev. X {\bf
  10},  021067  (2020).

\bibitem{dw}
M. Willsch, D. Willsch, F. Jin, H. De~Raedt, and K. Michielsen, Quantum
  Information Processing {\bf 19},  197  (2020).

\bibitem{1038}
E. Pelofske, G. Hahn, and H.~N. Djidjev, Scientific Reports {\bf 12},  4499
  (2022).

\bibitem{cit222y}
E. Pelofske, G. Hahn, and H.~N. Djidjev, Quantum Information Processing {\bf
  22},  219  (2023).

\bibitem{Niu2023enablingmulti}
S. Niu and A. Todri-Sanial, {Quantum} {\bf 7},  925  (2023).

\bibitem{3352460.3358287}
P. Das, S.~S. Tannu, P.~J. Nair, and M. Qureshi,  in {\em Proceedings of the
  52nd Annual IEEE/ACM International Symposium on Microarchitecture}, {\em
  MICRO '52} (Association for Computing Machinery, New York, NY, USA, 2019),
  p.\ 291–303.

\bibitem{Mineh_2023}
L. Mineh and A. Montanaro, Quantum Science and Technology {\bf 8},  035012
  (2023).

\bibitem{9749894}
Y. Ohkura, T. Satoh, and R. Van~Meter, IEEE Transactions on Quantum Engineering
  {\bf 3},  1  (2022).

\bibitem{10.00005}
A. Lucas, Frontiers in Physics {\bf 2},    (2014).

\bibitem{lodewijks2020mapping}
B. Lodewijks, Mapping NP-hard and NP-complete optimisation problems to
  Quadratic Unconstrained Binary Optimisation problems, arXiv:~1911.08043,
  2020.

\bibitem{Raeisi_2012}
S. Raeisi, N. Wiebe, and B.~C. Sanders, New Journal of Physics {\bf 14},
  103017  (2012).

\bibitem{042303}
D. Wecker, M.~B. Hastings, and M. Troyer, Phys. Rev. A {\bf 92},  042303
  (2015).

\bibitem{gokhale2019minimizing}
P. Gokhale, O. Angiuli, Y. Ding, K. Gui, T. Tomesh, M. Suchara, M. Martonosi,
  and F.~T. Chong, Minimizing State Preparations in Variational Quantum
  Eigensolver by Partitioning into Commuting Families, arXiv:~1907.13623, 2019.

\bibitem{venturelli2016quantum}
D. Venturelli, D.~J.~J. Marchand, and G. Rojo, Quantum Annealing Implementation
  of Job-Shop Scheduling, arXiv:~1506.08479, 2016.

\bibitem{90.022305}
D. Wecker, B. Bauer, B.~K. Clark, M.~B. Hastings, and M. Troyer, Phys. Rev. A
  {\bf 90},  022305  (2014).

\bibitem{jena2019pauli}
A. Jena, S. Genin, and M. Mosca, Pauli Partitioning with Respect to Gate Sets,
  arXiv:~1907.07859, 2019.

\bibitem{10.1063}
V. Verteletskyi, T.-C. Yen, and A.~F. Izmaylov, The Journal of Chemical Physics
  {\bf 152},    (2020), 124114.

\bibitem{rannea1}
A. Perdomo-Ortiz, S.~E. Venegas-Andraca, and A. Aspuru-Guzik, Quantum
  Information Processing {\bf 10},  33  (2011).

\bibitem{022314}
M. Ohkuwa, H. Nishimori, and D.~A. Lidar, Phys. Rev. A {\bf 98},  022314
  (2018).

\bibitem{arxiv.2303.13748}
E. Pelofske, G. Hahn, and H. Djidjev, Initial state encoding via reverse
  quantum annealing and h-gain features, arXiv:~2303.13748, 2023.

\bibitem{Venturelli_2019}
D. Venturelli and A. Kondratyev, Quantum Machine Intelligence {\bf 1},  17
  (2019).

\bibitem{PhysRevA.100.052321}
Y. Yamashiro, M. Ohkuwa, H. Nishimori, and D.~A. Lidar, Phys. Rev. A {\bf 100},
   052321  (2019).

\bibitem{izmaylov2019unitary}
A.~F. Izmaylov, T.-C. Yen, R.~A. Lang, and V. Verteletskyi, Unitary
  partitioning approach to the measurement problem in the Variational Quantum
  Eigensolver method, arXiv:~1907.09040, 2019.

\bibitem{cite111key}
C. Silva, A. Aguiar, P.~M.~V. Lima, and I. Dutra, Quantum Machine Intelligence
  {\bf 2},  16  (2020).

\bibitem{kwok2020graph}
J. Kwok and K. Pudenz, Graph Coloring with Quantum Annealing,
  arXiv:~2012.04470, 2020.

\bibitem{Kole2022}
A. Kole, D. De, and A.~J. Pal,  in {\em Intelligence Enabled Research: DoSIER
  2021}, edited by S. Bhattacharyya, G. Das, and S. De (Springer Singapore,
  Singapore, 2022), pp.\ 1--15.

\bibitem{PhysRevApplied.19.024047}
M.~S. Jattana, F. Jin, H. De~Raedt, and K. Michielsen, Phys. Rev. Appl. {\bf
  19},  024047  (2023).

\bibitem{willsch2022hybrid}
D. Willsch, M. Jattana, M. Willsch, S. Schulz, F. Jin, H.~D. Raedt, and K.
  Michielsen,  in {\em Hybrid Quantum Classical Simulations}, Vol.~51 of {\em
  Publication Series of the John von Neumann Institute for Computing (NIC) NIC
  Series}, NIC-Symposium, Germany, Jülich, 29 Sep 2022 - 30 Sep 2022, edited
  by M. Müller, C. Peter, and A. Trautmann (Forschungszentrum Jülich GmbH
  Zentralbibliothek, Verlag, Jülich, 2022), p.\ 450.

\bibitem{jattana}
M.~S. Jattana, {\em Applications of variational methods for quantum computers}
  (RWTH Aachen University, Aachen, 2022).

\bibitem{44}
A. Kandala, A. Mezzacapo, K. Temme, M. Takita, M. Brink, J.~M. Chow, and J.~M.
  Gambetta, Nature {\bf 549},  242  (2017).

\bibitem{45}
Y. Nam, J.-S. Chen, N.~C. Pisenti, K. Wright, C. Delaney, D. Maslov, K.~R.
  Brown, S. Allen, J.~M. Amini, J. Apisdorf, K.~M. Beck, A. Blinov, V. Chaplin,
  M. Chmielewski, C. Collins, S. Debnath, K.~M. Hudek, A.~M. Ducore, M. Keesan,
  S.~M. Kreikemeier, J. Mizrahi, P. Solomon, M. Williams, J.~D. Wong-Campos, D.
  Moehring, C. Monroe, and J. Kim, npj Quantum Information {\bf 6},  33
  (2020).

\bibitem{46}
C. Kokail, C. Maier, R. van Bijnen, T. Brydges, M.~K. Joshi, P. Jurcevic, C.~A.
  Muschik, P. Silvi, R. Blatt, C.~F. Roos, and P. Zoller, Nature {\bf 569},
  355  (2019).

\bibitem{kp1972}
R.~M. Karp,  in {\em Complexity of Computer Computations: Proceedings of a
  symposium on the Complexity of Computer Computations, held March 20--22,
  1972, at the IBM Thomas J. Watson Research Center, Yorktown Heights, New
  York, and sponsored by the Office of Naval Research, Mathematics Program, IBM
  World Trade Corporation, and the IBM Research Mathematical Sciences
  Department}, edited by R.~E. Miller, J.~W. Thatcher, and J.~D. Bohlinger
  (Springer US, Boston, MA, 1972), pp.\ 85--103.

\bibitem{glover}
F. Glover, G. Kochenberger, R. Hennig, and Y. Du, Annals of Operations Research
  {\bf 314},  141  (2022).

\bibitem{Chancellor_2019}
N. Chancellor, Quantum Science and Technology {\bf 4},  045004  (2019).

\bibitem{9259934}
Z. Tabi, K.~H. El-Safty, Z. Kallus, P. Haga, T. Kozsik, A. Glos, and Z.
  Zimboras,  in {\em 2020 IEEE International Conference on Quantum Computing
  and Engineering (QCE)} (IEEE Computer Society, Los Alamitos, CA, USA, 2020),
  pp.\ 56--62.

\bibitem{qubogen}
Y. Tamura and M. Sakai, \url{https://github.com/tamuhey/qubogen}, accessed:
  2023-05-05.

\bibitem{SciPyProceedings_11}
A. Hagberg, P.~J. Swart, and D.~A. Schult,  in {\em Exploring network
  structure, dynamics, and function using NetworkX} (USDOE, United States,
  2008).

\bibitem{dwave}
D-Wave Leap, \url{https://www.dwavesys.com/take-leap}, accessed: 2023-05-05.

\bibitem{PhysRevX.6.031007}
P.~J.~J. O'Malley, R. Babbush, I.~D. Kivlichan, J. Romero, J.~R. McClean, R.
  Barends, J. Kelly, P. Roushan, A. Tranter, N. Ding, B. Campbell, Y. Chen, Z.
  Chen, B. Chiaro, A. Dunsworth, A.~G. Fowler, E. Jeffrey, E. Lucero, A.
  Megrant, J.~Y. Mutus, M. Neeley, C. Neill, C. Quintana, D. Sank, A.
  Vainsencher, J. Wenner, T.~C. White, P.~V. Coveney, P.~J. Love, H. Neven, A.
  Aspuru-Guzik, and J.~M. Martinis, Phys. Rev. X {\bf 6},  031007  (2016).

\bibitem{dev:jakarta}
{7-qubit backend: IBM Q team, "IBM Q 7 Jakarta backend specification V1.2.7,"
  (2023)}.

\bibitem{dev:manila}
{5-qubit backend: IBM Q team, "IBM Q 5 Manila backend specification V1.1.4,"
  (2023)}.

\bibitem{dev:perth}
{7-qubit backend: IBM Q team, "IBM Q 7 Perth backend specification V1.1.43,"
  (2023)}.

\bibitem{dev:nairobi}
{7-qubit backend: IBM Q team, "IBM Q 7 Nairobi backend specification V1.2.6,"
  (2023)}.

\bibitem{fphy.2022.907160}
M.~S. Jattana, F. Jin, H. De~Raedt, and K. Michielsen, Frontiers in Physics
  {\bf 10},    (2022).

\bibitem{McClean_2020}
J.~R. McClean, N.~C. Rubin, K.~J. Sung, I.~D. Kivlichan, X. Bonet-Monroig, Y.
  Cao, C. Dai, E.~S. Fried, C. Gidney, B. Gimby, P. Gokhale, T. Häner, T.
  Hardikar, V. Havlíček, O. Higgott, C. Huang, J. Izaac, Z. Jiang, X. Liu, S.
  McArdle, M. Neeley, T. O’Brien, B. O’Gorman, I. Ozfidan, M.~D. Radin, J.
  Romero, N.~P.~D. Sawaya, B. Senjean, K. Setia, S. Sim, D.~S. Steiger, M.
  Steudtner, Q. Sun, W. Sun, D. Wang, F. Zhang, and R. Babbush, Quantum Science
  and Technology {\bf 5},  034014  (2020).

\bibitem{PhysRevX.8.031022}
C. Hempel, C. Maier, J. Romero, J. McClean, T. Monz, H. Shen, P. Jurcevic,
  B.~P. Lanyon, P. Love, R. Babbush, A. Aspuru-Guzik, R. Blatt, and C.~F. Roos,
  Phys. Rev. X {\bf 8},  031022  (2018).

\bibitem{1063}
J.~T. Seeley, M.~J. Richard, and P.~J. Love, The Journal of Chemical Physics
  {\bf 137},    (2012), 224109.

\bibitem{10.10931.85}
D.~J.~A. Welsh and M.~B. Powell, The Computer Journal {\bf 10},  85  (1967).

\bibitem{6.013205}
C. Cao, H. Yano, and Y.~O. Nakagawa, Phys. Rev. Res. {\bf 6},  013205  (2024).

\bibitem{Cao_2023}
C. Cao, Y. Yu, Z. Wu, N. Shannon, B. Zeng, and R. Joynt, Quantum Science and
  Technology {\bf 8},  015004  (2022).

\bibitem{citekeyem}
M.~S. Jattana, F. Jin, H. De~Raedt, and K. Michielsen, Quantum Information
  Processing {\bf 19},  414  (2020).

\bibitem{909654}
T. Lippert and K. Michielsen,  in {\em Perspectives of Quantum Computing at the
  Jülich Supercomputing Centre}, Vol.~51 of {\em Publication Series of the
  John von Neumann Institute for Computing (NIC) NIC Series}, NIC-Symposium,
  Germany, Jülich, 29 Sep 2022 - 30 Sep 2022, edited by M. Müller, C. Peter,
  and A. Trautmann (Forschungszentrum Jülich GmbH Zentralbibliothek, Verlag,
  Jülich, 2022), p.\ 450.

\end{thebibliography}
\end{document}